\documentclass[preprint]{aastex}

\usepackage{graphicx}

\DeclareMathAlphabet{\mathbf}{OT1}{cmr}{bx}{it}

\begin{document}

\title{Polarimetric Diagnostics of Unresolved Chromospheric Magnetic 
Fields}

\author{R. Casini,$^a$ R. Manso Sainz,$^b$ B. C. Low$^a$}

\affil{$^a\,$High Altitude Observatory,
National Center for Atmospheric Research,\altaffilmark{1}
P.O.~Box 3000, Boulder, CO 80307-3000\\
$^b\,$Instituto de Astrof\'{\i}sica de Canarias, c/ V\'{\i}a 
L\'actea s/n, La Laguna, Tenerife, E-38200 Spain}

\altaffiltext{1}{The National Center for Atmospheric Research is
sponsored by the National Science Foundation.}

\begin{abstract}
For about a decade, spectro-polarimetry of \ion{He}{1} $\lambda$10830 
has been applied to the magnetic diagnostics of the solar chromosphere. 
This resonance line is very versatile, as it is visible both on disk 
and in off-limb structures, and it has a good sensitivity to both the 
weak-field Hanle effect and the strong-field Zeeman effect. Recent 
observations of an active-region filament showed that the linear 
polarization was dominated by the transverse Zeeman effect, with very 
little or no hint of scattering polarization. This is surprising, 
since the \ion{He}{1} levels should be significantly polarized
in a conventional scattering scenario.
To explain the observed level of atomic depolarization by collisional 
or radiative processes, one must invoke plasma densities larger by 
several orders of magnitude 
than currently known values for prominences. We show that
such depolarization can be explained quite naturally by the presence 
of an unresolved, highly entangled magnetic field, which averages to
give the ordered field inferred from spectro-polarimetric data, over the 
typical temporal and spatial scales of the observations. We present 
a modeling of the polarized \ion{He}{1} $\lambda$10830 in this 
scenario, and discuss its implications for the magnetic diagnostics 
of prominences and spicules, and for the general study 
of unresolved magnetic field distributions in the solar atmosphere.
\end{abstract}

\subjectheadings{Sun: chromosphere -- Sun: prominences -- Sun: magnetic
fields -- line: profiles -- polarimetry}

\maketitle

Magnetic diagnostics of the chromosphere, and in particular 
of prominences and spicules, is receiving increasing attention and 
motivation from the solar community. There is in fact a growing 
agreement that mapping the magnetic field in these critical 
regions of the solar atmosphere is fundamental for our understanding 
of processes like coronal heating, the acceleration of the solar 
wind, and the release of coronal mass ejections, that have 
a direct influence on the heliosphere and on the associated phenomena 
of space weather. Despite the strong demand for these critical
observations, there have been relatively few attempts to 
measure magnetic fields in the chromosphere and corona. This is 
mainly due to the heavy science requirements imposed on 
spectro-polarimetric instrumentation by this type of observations, and
to the intrinsic difficulty of the inversion and interpretation of
scattering polarization data.

Spectral line polarization is produced when symmetry-breaking 
processes occur in the interaction of radiation with
matter. These can be due to the presence of external magnetic or
electric fields (Zeeman and Stark effects), or an anisotropy in the
excitation of the atoms (by radiation or particles) leading to scattering 
polarization. We can expect both types of processes to be always 
present in the solar atmosphere. However, their corresponding
polarization signals differ significantly. In particular, the presence 
of unresolved fields affects those signals in characteristic ways. 
For example, for a completely random distribution of magnetic fields
within the resolution element of the observations, the 
polarization by the Zeeman effect must vanish in the mean, whereas 
the scattering polarization is only reduced in amplitude by a 
characteristic factor of 1/5 with respect to the zero-field case. 
When both processes occur, and the random magnetic field has a 
non-zero mean, the observed signal will be a complex mix 
of Zeeman effect and scattering polarization, with a significant
depolarization contributed by the random part of the magnetic field.

Recent observations of the \ion{He}{1} multiplet at 1083\,nm in an 
active-region (AR) filament \citep{Ku09} have shown that the 
linear polarization was dominated by the transverse Zeeman effect, 
corresponding to magnetic fields in the 500\,G--1000\,G range. 
The modeling of the forward-scattered radiation of \ion{He}{1} 
$\lambda$10830, under the assumption that the filament plasma is 
illuminated by the underlying photosphere, indicates instead that 
the linear polarization should be significantly affected by the 
atomic alignment of the upper term $\rm 2\,{}^3P$, induced by 
radiation anisotropy, even in the presence of field strengths of 
the order of $10^3$\,G.
A basic assumption of the Zeeman-effect model is that the atomic 
levels of \ion{He}{1} are ``naturally'' populated, and therefore 
completely depolarized. In contrast, in the scattering-polarization
model, which includes atomic polarization, we were forced to 
introduce an ad-hoc reduction factor for the radiation anisotropy 
(varying between 0 and 1) in order to reproduce the observed 
depolarization of the \ion{He}{1} levels. Obviously both 
models lack a physical basis for such depolarization, which is 
the question taken up in this Letter.

The characteristic rate, $\gamma$, at which atomic polarization 
is generated in the levels of a spectral line by resonance 
scattering, is inversely proportional to the radiative lifetimes of 
the levels. In particular, for \ion{He}{1} $\lambda$10830,
$\gamma_u\sim A_{ul}\approx 10^7\rm\,s^{-1}$ and $\gamma_l\sim B_{lu}J$, 
where $A_{ul}$ and $B_{lu}$ are the Einstein coefficients, 
respectively for spontaneous emission and absorption, between the upper 
($u$) and lower ($l$) levels, and $J$ is the average intensity of 
the radiation field. If we assume
$J\approx (1/2)\,B_\nu(T{=}5800\,\rm K)$, then
$\gamma_l\approx(3/20)\,\gamma_u$. 

Once created, atomic polarization is not easily destroyed, 
unless competing processes (e.g., external fields, collisions, radiative 
ionization and recombination) modify the polarization of the 
atomic levels at a much higher rate than both $\gamma_l$ and 
$\gamma_u$. A strong magnetic field (i.e., with Larmor frequency
$\nu_{\rm L}\gg\gamma_{l,u}$) can depolarize 
atomic levels (through the Hanle effect) very efficiently, 
although the degree of
residual polarization strongly depends on the geometry of the field.
For example, scattering polarization is completely destroyed by 
a strong magnetic field with inclination 
$\vartheta_B=\arccos1/\sqrt{3}\approx 54.7^\circ$
from the direction of illumination. This effect may be invoked 
to explain the AR filament observations described in \cite{Ku09}. 
However, those authors report profiles showing strong atomic 
depolarization even for nearly horizontal magnetic fields.

\begin{figure}[!t]
\centering
\includegraphics[height=3in]{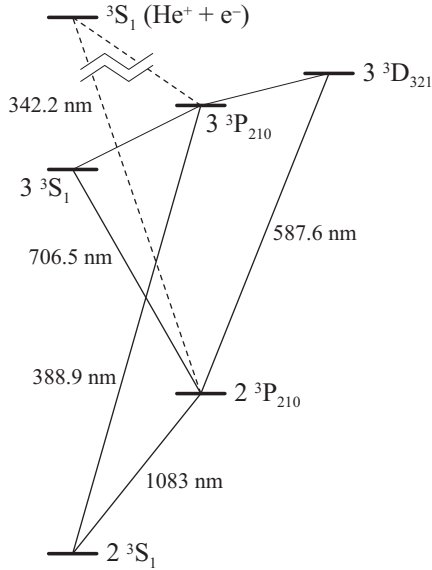}
\caption{\label{fig:grotrian}
Model atom for the statistical equilibrium of the triplet
species of \ion{He}{1}. The lowest five terms constitute the system of
bound-bound transitions. The model is extended to include 
photo-ionization and recombination involving the $\rm {}^3P$ states, 
by adding a fictitious bound state $\rm {}^3S_1$ corresponding to 
the ground state of \ion{He}{2}.}
\end{figure}

Atomic collisions with an isotropic distribution of perturbers
(neutral or charged) are a possible mechanism of atomic
depolarization. 1) Inelastic collisions with electrons are not
significant for the statistical equilibrium of \ion{He}{1} at
the typical densities of prominences 
($N\sim10^{10}$--$10^{11}\,\rm cm^{-3}$). 
The observation of scattering polarization at the limb in these 
structures also points to the fact that atomic excitation is dominated 
by resonance scattering rather than thermal processes. 
It is possible 
that AR filaments may be significantly denser than quiescent prominences,
leading to a larger contribution of thermal processes. However,
following \cite{vR62}, collisional excitation of 
\ion{He}{1} $\lambda$10830 by electrons would require 
$n_{\rm e}\approx 4\times10^{13}\,\rm cm^{-3}$, in order to be comparable 
with radiative excitation. This would imply gas densities 3 to 4 
orders of magnitude larger than currently known values in quiescent 
prominences \citep[e.g.,][]{TH95} for
electron collisions to dominate the formation of \ion{He}{1}.
2) We can estimate the rate of elastic 
collisions of \ion{He}{1} atoms in the $\rm {}^3P$ state with 
neutral hydrogen, using 
$\gamma_{\rm H}\sim\langle\sigma v\rangle\,n_{\rm H}\approx 8\times
10^{-9}\,n_{\rm H}\,\rm cm^3\,s^{-1}$ \citep{LT71}. 
Hence, to efficiently depolarize the upper levels of \ion{He}{1} 
$\lambda10830$ by collisions with neutral hydrogen it should be
$n_{\rm H}\gg 10^{15}\,\rm cm^{-3}$.
3) Elastic collisions with electrons have received little
attention in the literature. \cite{Hi88} report experimental
cross-sections at $T=2000$\,K for \ion{Ne}{1} in the $2p_2$ level.
If we extend their results to the case of \ion{He}{1} at
plasma temperatures $T\sim 10\,000$\,K, we find 
$\gamma_{\rm e}^{\rm el.}\sim 10^{-6}\,n_{\rm e}\,\rm cm^3\,s^{-1}$. 
Efficient depolarization of the upper state of \ion{He}{1} 
$\lambda10830$ would then imply $n_{\rm e}\gg 10^{13}\,\rm cm^{-3}$, 
at which line excitation by thermal electrons should 
dominate over radiative excitation (see above).
In conclusion, collisional depolarization of the \ion{He}{1} levels
seems to consistently require plasma densities that are exceptionally 
large for a typical prominence. On the other hand, the densities of 
AR filaments are poorly known, and the possibility of such high
plasma densities merits further investigation.

\begin{figure}[!t]
\centering
\includegraphics[width=\hsize]{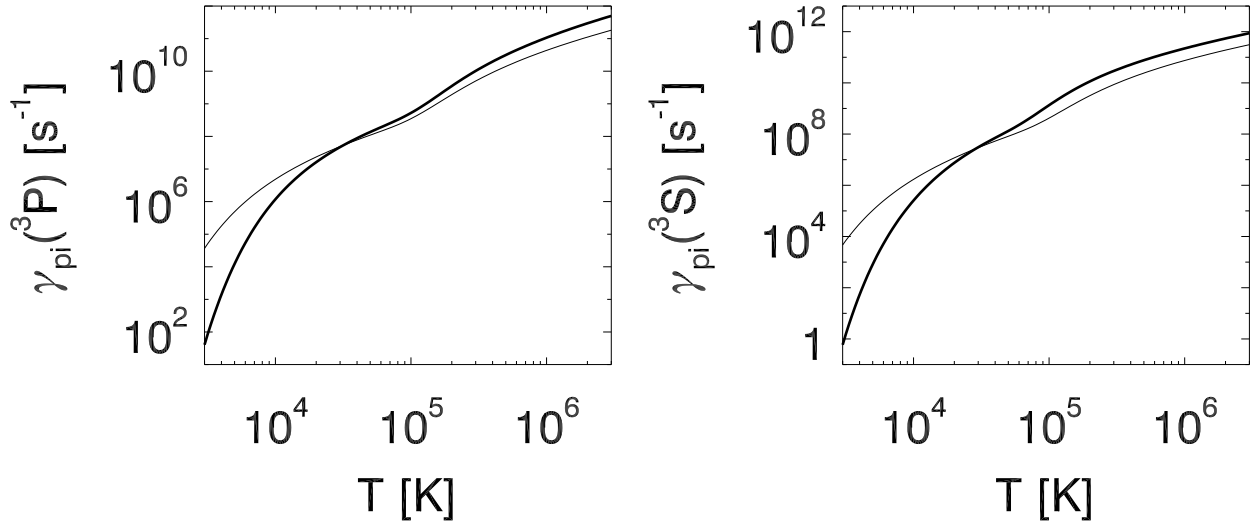}
\figcaption{\label{fig:Xsect}
Photo-ionization rates for the lowest $\rm {}^3P$ terms (left) and 
$\rm {}^3S$ terms (right) of neutral helium, as a function 
of the temperature of a Planckian illumination. The thick lines
correspond to the $\rm 2\,{}^3X$ terms, and the thin lines to the
$\rm 3\,{}^3X$ terms.}
\end{figure}

We consider next the role of radiative processes like 
photo-ionization and recombination for the depolarization of 
\ion{He}{1} levels.
Figure~\ref{fig:grotrian} illustrates the atomic model commonly 
employed for the study of resonance scattering polarization in 
\ion{He}{1} lines (e.g., \citealt{LA82,LC02,AT08}). This model has 
been successfully applied to interpret the Stokes profiles 
of the D$_3$ line at 587.6\,nm in solar prominences 
\citep{BS78,LA82,CA03}, and in more recent years to the 
magnetic diagnostics of quiescent prominences and filaments observed 
at 1083\,nm \citep{TB02,Me06}. 
The system of bound-bound transitions involves the 
lowest five terms of the triplet species of \ion{He}{1}. In order to 
include photo-ioniziation and recombination in this model, we added a 
fictitious bound state, $\rm {}^3S_1$, corresponding to the
ground level of \ion{He}{2} plus a free electron (with spin parallel to
that of the \ion{He}{2} ion). 
The depopulation rate by photo-ionization is given by
$\gamma_{\rm pi}=\int_{\chi/h}^\infty d\nu\;
	(4\pi a_\nu/h\nu)\,J(\nu)$,
where $J(\nu)$ is the intensity of the ionizing radiation 
averaged over the unit sphere, and $\chi$ is the ionization potential.
The ionization cross-section, $a_\nu$, for various atomic terms of
\ion{He}{1} is tabulated by the TOPbase project \citep[e.g.,][]{BA05}.
In the case of the $\rm 2\,{}^3P$ term, the frequency range over which
the cross-section is tabulated goes from the ionization threshold up 
to $\sim 3.7$\,nm.
Unfortunately, the mean intensity of the solar radiation below the 
ionization wavelength of 342.2\,nm is not well known.
UV observations of the solar atmosphere have traditionally focused on 
limited spectral ranges of particular interest. More complete
datasets, covering the whole UV spectrum down to 
$\lambda\sim 1$\,nm \citep{EW03}, are integrated over the solar
disk, and do not distinguish the output of the quiet Sun from that of 
active regions (although they account for variability of the irradiance
during the solar cycle).
Figure~\ref{fig:Xsect} shows the photo-ionization rate,
$\gamma_{\rm pi}$, of the \ion{He}{1} terms as a function of 
the effective (Planckian) temperature of the plasma contributing the
ionizing radiation. We see that 
$\gamma_{\rm pi}\sim\gamma_u$
at effective temperatures of $10^5$\,K or larger. On the other
hand, the effective temperature of the UV emitting solar plasma, as 
estimated by the integrated number of UV photons detected above the 
Earth's atmosphere \citep{EW03}, is smaller by at least one order of 
magnitude. Therefore the ionizing UV radiation cannot play a 
major role in the depolarization of the \ion{He}{1} $\lambda$10830 
levels.

The depolarizing mechanisms considered above seem unable to 
efficiently destroy atomic polarization generated by resonance 
scattering, unless one can accept a radical revision of plasma density
estimates in AR filaments. The possibility remains that in the 
AR filament described by \cite{Ku09} atomic polarization was not
generated at all. We considered whether the center-to-limb 
variation (CLV) of the solar atmosphere, which dominates the 
anisotropy of the illuminating radiation at low
heights, may be significantly different in an active region, because 
of enhanced lateral illumination coming from plages and/or a brighter 
transition region. However, the few measurements of the CLV in active
regions \citep[e.g.,][]{SC02} covering the near-infrared solar
spectrum do not seem to support this possibility.

\cite{TA07} have proposed that the radiation inside an optically 
thick filament could be nearly isotropic, and
therefore unable to induce atomic polarization.
By modeling the filament as a free-standing, plane-parallel slab
of optical thickness $\tau\sim 1$ and constant source function, $S$,
illuminated by the underlying photosphere with intensity $I_0$, 
they find that the radiation anisotropy inside the slab may be 
significantly reduced and even become negative (because the 
illumination is predominantly horizontal). In particular, 
when $S\approx I_0$, the anisotropy inside the filament vanishes.
However, if the formation of \ion{He}{1} $\lambda10830$
is dominated by scattering 
($\gamma_{\rm e}^{\rm inel.}/\gamma_u\sim 0.002$ for 
$n_{\rm e}\sim10^{11}\,\rm cm^{-3}$ typical of prominences),
the source function must be modeled accordingly. For example, 
simply assuming 
$S\approx\oint (\mathbf{d\Omega}/4\pi)\,p(\cos\Theta)\,I$,
where $p(\cos\Theta)$ is the Rayleigh phase function, the anisotropy
rises sharply from about 40\% of its optically thin limit close to 
the lower boundary, to well above the optically thin limit at the 
free upper boundary. Thus, the anisotropy inside the slab does fall 
below the optically thin limit, but complete isotropy is never 
attained. Therefore, although a careful radiative transfer treatment
of this problem is important, the atomic depolarization of \ion{He}{1} 
may not depend only on that. Moreover, the inversions 
presented by \cite{Ku09} indicated that the average optical depth of 
the filament was sizably smaller than 1, further reducing the 
depolarizing role of radiative transfer.

The preceeding analysis motivated us to consider a mechanism of 
depolarization not commonly included in polarimetric diagnostics, 
namely, that the depolarization of the
\ion{He}{1} levels is due to the presence of a significant random 
component in the magnetic field inferred from
spectro-polarimetric observations.  Both observation and theory support 
the idea that solar magnetic fields, even in the quasi-steady state, 
are chaotic and dynamic on small spatial and temporal scales.
The quiescent-prominence movies made by Hinode/SOT at unprecedented
high temporal and spatial resolutions have revealed the extreme 
complexity and rapid evolution of filamentary prominence plasma
\citep{Be08}.  The long lifetime of a quiescent prominence, days to 
weeks, implies a large-scale, stable magnetic topology, but which is 
separate from its significantly entangled
and rapidly evolving structures at subarsecond scales. MHD waves and 
oscillations, together with rising and descending
plumes, develop ceaselessly in the macroscopic prominence structure.
Although the AR filament described by
\cite{Ku09} is a class of prominence distinct from the quiescent 
prominences, we may expect the average field inferred from observation 
to be co-existing with a small-scale, random component of a comparable 
field intensity.  In the extremely low-$\beta$, electrically highly 
conducting environment of the AR atmosphere above the photosphere, 
current sheets are expected to form densely within any volume 
pervaded by a magnetic field endowed with a complex topology
\citep{Pa94,JL09}. As these sheets dissipate to heat the atmosphere 
quiescently, new sheets form without necessarily producing a major 
flare.  The small-scale and rapidly evolving fields associated with a 
complexity of current sheets are naturally of an intensity comparable 
to the mean field, and could be the origin of the quasi-random 
magnetic field we consider here.

\begin{figure}[!t]
\centering
\includegraphics[width=\hsize]{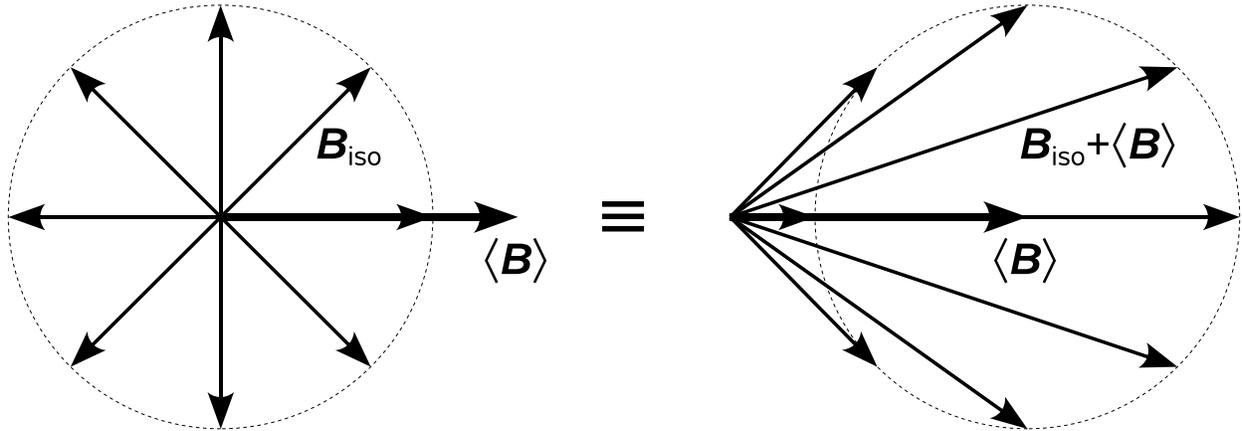}
\caption{\label{fig:model}
Quasi-random magnetic field model (right), obtained by
adding a perfectly isotropic (random) field, 
$\mathbf{B}_{\rm iso}$, to the inferred mean field, 
$\langle\mathbf{B}\rangle$ (left).}
\end{figure}

\begin{figure*}[!t]
\centering
\includegraphics[width=.495\hsize]{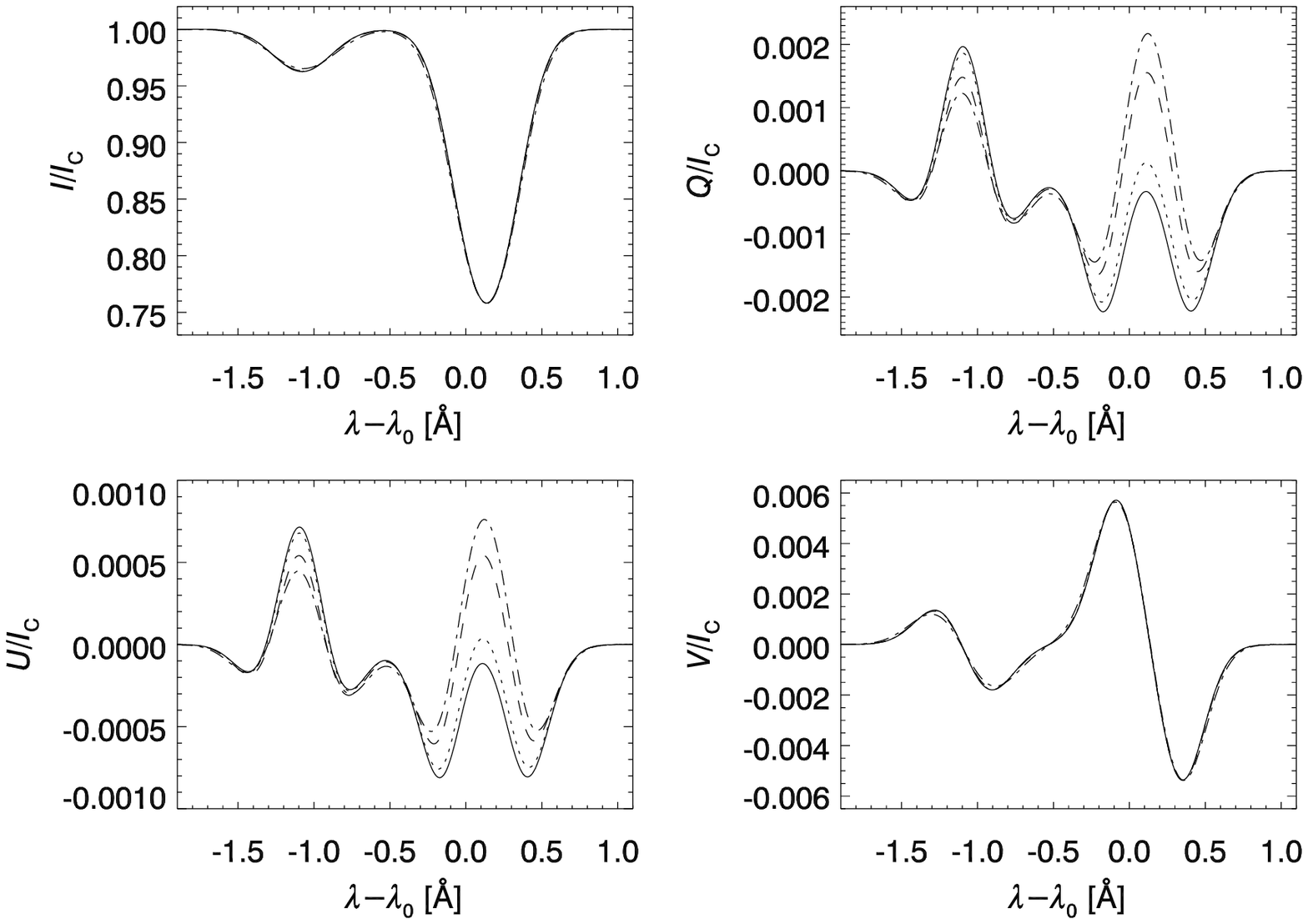}\kern 5pt
\includegraphics[width=.495\hsize]{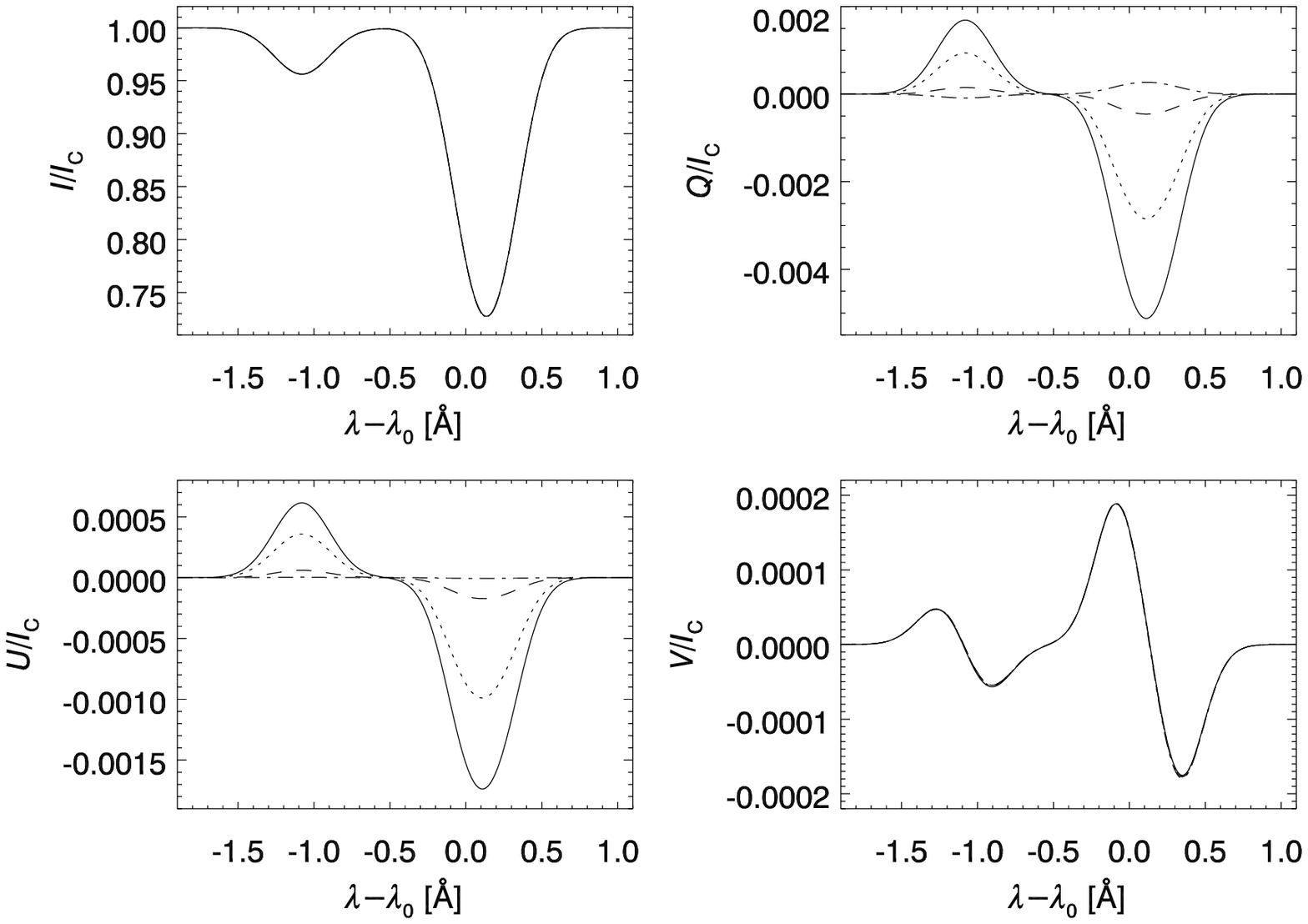}
\caption{\label{fig:Stokes}
Examples of Stokes profiles of \ion{He}{1} $\lambda$10830 observed 
on the disk, in the presence of a nearly horizontal magnetic field 
plus a completely isotropic field of various strengths. 
\textit{Left:} $\langle B\rangle=700$\,G, 
$B_{\rm iso}=0$\,G (continuous curve), 
200\,G (dotted curve), 
500\,G (dashed curve), 
1000\,G (dashed-dotted curve). \textit{Right:}
$\langle B\rangle=20$\,G,
$B_{\rm iso}=0$\,G (continuous curve), 
10\,G (dotted curve), 
20\,G (dashed curve), 
100\,G (dashed-dotted curve).}
\end{figure*}

Following this picture, we have calculated the emergent Stokes
profiles from a homogeneous slab, assuming that the \ion{He}{1} atoms
are subject to the mean field inferred from the observations plus a 
completely random (hence, isotropic) field of a given strength. 
This is also equivalent to assuming a specific 
non-isotropic distribution of magnetic fields with varying strength, 
resulting exactly in the observed mean field (Fig.~\ref{fig:model}). 
It can be argued that this model is very likely an oversimplification 
of the real, quasi-random field occurring in prominences. However,
its main purpose is to illustrate in the simplest way the depolarizing 
effect of such a field. There is no difficulty in principle in calculating 
the same effect also for more complicated field distributions.
Figure~\ref{fig:Stokes} (left) shows the emergent Stokes profiles for such a 
distribution of magnetic fields, resulting in an almost horizontal 
mean field of 700\,G, similar to those described by \cite{Ku09}. 
The four overplotted profiles illustrate the depolarizing effect induced 
by the inclusion of an isotropic field of 200, 500, and 1000\,G, as well 
as the case where only the mean field is present. The level of atomic 
depolarization inferred from the observations corresponds to an 
isotropic field between 500 and 1000\,G, in agreement with the 
proposed picture of a very entangled, strong field occurring in the 
filament plasma.
Figure~\ref{fig:Stokes} (right) also shows the expected effects 
of atomic depolarization in the presence of a quasi-random field 
averaging to a nearly horizontal, ordered magnetic field with 
a strength of 20\,G, typical of quiescent prominences and 
spicules. Especially at such small field strengths, the inference
of the magnetic field by the Hanle effect becomes very sensitive to 
the presence of additional atomic depolarizing processes.

The successful explanation of atomic depolarization by a quasi-random
magnetic field stresses the importance to include unresolved fields in
the polarimetric diagnostic investigation of the solar atmosphere. The 
depolarization effect 
of a completely turbulent magnetic field has been studied in detail 
\citep[e.g.,][]{St94,LL04}, and has also found many applications in the 
study of 
quiet-Sun magnetism \citep[see, e.g., the review by][]{TB06}. 
In contrast, to our knowledge, no general account has been given in
the past of similar depolarization effects produced by a field that 
is significantly entangled at scales below the temporal and spatial 
resolutions of spectro-polarimetric observations, but which averages 
at a finite, mean field when integrated over those scales.
Such mean field, which determines the large-scale evolution of the 
plasma, remains within the grasp of the traditional diagnostics based 
on the Zeeman and Hanle effects. On the other hand, the signature of 
atomic depolarization---like in the observations described by 
\cite{Ku09}---finds a very simple and direct interpretation in 
terms of the strength of a completely random field unresolved 
to the observations. This scenario provides a very appealing 
interpretational framework for the highly dynamical events 
observed at high temporal and spatial resolution with Hinode/SOT, 
e.g., in quiescent prominences and spicules \citep{Be08}, thus 
opening a new path for the polarimetric study of highly structured 
plasmas. 

Of course, this new insight comes at the cost of introducing 
an additional degree of freedom in the magnetic diagnostics of 
chromospheric fields. For this reason, we strongly advocate that
future research programs need to take full advantage of multi-line 
spectro-polarimetry, e.g., through simultaneous observations of 
\ion{He}{1} $\lambda$10830 and D$_3$ in prominences and spicules,
or \ion{He}{1} $\lambda$10830 and H$\alpha$ $\lambda$6563 in 
filaments and the chromosphere. In fact, such capability is already 
been pursued at the French-Italian solar telescope TH\'EMIS, and 
will be the normal mode of operation of HAO's Promincence Magnetometer 
\citep[ProMag;][]{El08}, soon to be deployed. We expect that this 
type of multi-line diagnostics of chromospheric fields will fully 
come to fruition with the large solar facilities of the next 
generation, like ATST, EST, and COSMO.

\acknowledgments We thank R.~Centeno Elliott (HAO) and
V.~Mart\'{\i}nex Pillet (IAC) for useful comments. RMS acknowledges 
support by the Spanish MCYT through project AYA2007-63881.

\end{document}